\documentclass[12pt,preprint]{aastex}
\def\chandra{{\it Chandra}}

\def\suzaku{{\it Suzaku}}

\def\asca{{\it ASCA}}

\newcommand{\Msun}{\hbox{$\rm\thinspace M_{\odot}$}}
\newcommand{\ls}
{\mathrel{\hbox{\rlap{\hbox{\lower4pt\hbox{$\sim$}}}\hbox{$<$}}}}
\newcommand{\gs}
{\mathrel{\hbox{\rlap{\hbox{\lower4pt\hbox{$\sim$}}}\hbox{$>$}}}}
\def\Msun{\hbox{$\rm ~M_{\odot}$}}
\def\3c{3C~382}

\begin{document}

\title{Chandra detection of a parsec scale wind in the Broad Line 
Radio Galaxy 3C\,382}

\author{J. N. Reeves}
\affil{Astrophysics Group, School of Physical \& Geographical Sciences, Keele
University, Keele, Staffordshire ST5 5BG, UK; e-mail jnr@astro.keele.ac.uk}

\author{R. M. Sambruna}
\affil{NASA/GSFC, Code 662, Greenbelt, MD 20771, USA}

\author{V. Braito}
\affil{University of Leicester, Department of Physics \& Astronomy,
University Road, Leicester LE1 7RH, UK}

\author{Michael Eracleous}
\affil{Department of Astronomy \& Astrophysics and Center for Gravitational Wave 
Physics, The Pennsylvania State University, 
525 Davey Laboratory, University Park, PA 16802, USA}

\begin{abstract}

We present unambiguous evidence for a parsec scale wind in
the Broad-Line Radio Galaxy (BLRG) 3C~382, the first radio-loud AGN, 
with $R_{\rm L} = \log_{10}(f_{\rm 5GHz}/f_{4400})>1$, 
whereby an outflow has been measured with X-ray grating spectroscopy.  A 118\,ks
Chandra grating (HETG) observation of 3C\,382 has revealed the
presence of several high ionization absorption lines in the soft X-ray
band, from Fe, Ne, Mg and Si. The absorption lines are blue-shifted
with respect to the systemic velocity of 3C\,382 by
$-840\pm60$\,km\,s$^{-1}$ and are resolved by Chandra with a velocity
width of $\sigma=340\pm70$\,km\,s$^{-1}$.  The outflow appears to originate from
a single zone of gas of column density $N_{\rm H} = 1.3 \times
10^{21}$\,cm$^{-2}$ and ionization parameter 
$\log (\xi / {\rm erg}\,{\rm cm}\,{\rm s}^{-1}) = 2.45$.
From the above measurements we calculate
that the outflow is observed on parsec scales, within the likely range
from $10-1000$\,pc, i.e., consistent with an origin in the Narrow Line
Region.  

\end{abstract}


\section{Introduction}

At least 50\% of radio-quiet AGN exhibit evidence for photoionized
outflows in their X-ray spectra \citep{Reynolds97, George98, Crenshaw03, 
Porquet04, Blustin05, McKernan07}. The signatures of these winds
consist of absorption and emission features at soft (0.4--2~keV) and
hard (6--8~keV) X-ray energies, coincident with ionized O, N, Ne, Mg,
Si and Fe lines blueshifted in the observer's rest-frame.  The
inferred velocities of the winds are typically in the range
$100-1000$\,km\,s$^{-1}$, but can be as high as $\sim 0.1c$ in some
sources \citep{Chartas02, Chartas03, Pounds03, Reeves03, Reeves08}. It is also
possible that the energy budget of the outflows of some AGN can
approach a significant fraction of the bolometric or even Eddington
luminosity \citep{KP03}.

In stark contrast, the X-ray evidence for nuclear outflows is very scarce 
in Broad-Line Radio Galaxies (BLRGs) and in radio-loud AGN generally. 
Previously, the radio-loud quasar 4C~+74.26 showed weak
absorption features at $\sim$ 1~keV with \asca\ \citep{Ballantyne05}, 
while the BLRG Arp\,102B showed neutral X-ray absorption with \asca\ and a 
UV outflow of a few hundred km\,s$^{-1}$ \citep{Erac03}.
Furthermore two radio galaxies, 3C~445 and 3C~33 \citep{Sambruna07, Evans06} 
also exhibit soft X-ray emission lines below 2~keV,
which could originate from spatially extended material (in
3C~33; \citet{Torresi09a}). Disk winds are also expected in
radio-loud AGN as ingredients for jet formation \citep{Blandford82}. 

To differentiate between radio-loud and radio-quiet AGN, here we adopt 
the radio-loudness parameter $R_{\rm L} = \log_{10}(f_{\rm 5GHz}/f_{4400})$, 
where $f_{\rm 5GHz}$ is the core 5\,GHz radio flux and $f_{4400}$ is the 
flux at 4400\AA, both in units of mJy \citep{Kellerman89}. Generally, 
$R_{\rm L}>1$ for radio-loud AGN \citep{Wilkes87}, while for 3C\,382, $R_{\rm L}=1.9$ 
\citep{Lawson97}. If the extended radio emission from 3C\,382 
is also included, then $R_{\rm L}$ may be considerably higher.

In this Letter we present direct evidence for outflowing gas from the nucleus
of the nearby ($z=0.05787$), bright BLRG \3c. A re-analysis of our
118~ks \chandra\ HETG (High Energy Transmission Grating) observations
\citep{Gliozzi07} revealed several blue-shifted absorption lines between
$0.7-2.0$\,keV which suggest the presence of a large-scale
($10-1000$\,pc) outflow in this source with a velocity of
800\,km\,s$^{-1}$.  
The organization of this Letter is as follows. In \S~2 we describe the
\chandra\ data reduction and analysis; in \S~3 the
results of the spectral analysis; Discussion and Conclusions follow in
\S~4. Throughout this paper, a concordance cosmology with H$_0=71$ km
s$^{-1}$ Mpc$^{-1}$, $\Omega_{\Lambda}$=0.73, and $\Omega_m$=0.27
\citep{Spergel03} is adopted. Errors are quoted to 90\% confidence 
for 1 parameter of interest (i.e. $\Delta \chi^{2}$ or $\Delta C=2.71$).

\section{The Chandra HETG Data} 


Chandra observed 3C\,382 with the HETG for a net exposure of 118\,ks
between 27--30 November 2005.  The $\pm1$ order spectra were summed
for the MEG (Medium Energy Grating) and HEG (High Energy Grating)
respectively, along with their response files.  The summed first order
count rates for the MEG and HEG are 0.867\,counts\,s$^{-1}$ and
0.379\,counts\,s$^{-1}$ respectively, while the MEG data were fitted
between 0.5--7.0\,keV and the HEG from 1.0--9.0\,keV.

\section{The Warm Absorber in 3C\,382}

The \chandra\ HETG data were first analyzed by \citet{Gliozzi07},
who focused on the continuum and its variability. Our results are in
agreement with theirs. Specifically, the MEG and HEG data were fitted
by an absorbed power law with photon index $\Gamma=1.66\pm0.01$ plus a
blackbody with $kT=92\pm6$~eV to parameterize the soft excess below
1~keV \citep{Gliozzi07}, absorbed by a Galactic line of sight
column of $N_{\rm H, Gal}=7.0\times10^{20}$\,cm$^{-2}$ \citep{Dickey90}.  
Figure~1 shows the broad-band HETG spectrum fitted
with an absorbed power-law only, to illustrate that a soft excess is clearly present
below 1\,keV.  The 0.5-9\,keV band flux is
$6.4\times10^{-11}$\,erg\,cm$^{-2}$\,s$^{-1}$.  Even upon adding a
blackbody to parameterize the soft excess, the fit is still formally
unacceptable ($\chi^2$/dof = 632/477, null probability $2.4 \times
10^{-6}$) as there are clear residuals around 1 keV that indicate the
presence of a warm absorber.

To analyse the warm absorber in detail, the HEG and MEG spectra were
binned more finely to sample the resolution of the detector, at
approximately HWHM the spectral resolution (e.g. $\Delta\lambda
=0.01$\AA\ bins for the MEG). For the fits, the C-statistic was used
\citep{Cash79}, as there are fewer than 20 counts per resolution bin. The
absorption lines were modelled with Gaussian profiles and the continuum model
was adopted from above.  Table~1
lists the detected lines with their observed and inferred properties,
and their significance as per the C-statistic.  Figure~2
shows the portions of the HETG spectrum containing the strongest
lines, with the model overlaid.

The seven absorption lines in Table~1 and Figure~2 are all detected at
high confidence (corresponding to $\Delta C>18$, or $>99.9\%$
confidence for 2 parameters of interest).  
The lines likely arise from the $1s-2p$ transitions of
Ne\,\textsc{ix}, Ne\,\textsc{x}, Si\,\textsc{xiii}, and
Si\,\textsc{xiv} and the $2p-3d$ lines of Fe\,\textsc{xix-xxi}. The
two statistically weaker $1s-2p$ lines of Mg\,\textsc{xi} and
Mg\,\textsc{xii} may also be present, which have outflow velocities 
consistent with the other lines.

Initially we assume that the lines have the same velocity width within
the errors. The velocity width of the absorption lines is then
$\sigma=340\pm70$\,km\,s$^{-1}$ (or $780\pm160$\,km\,s$^{-1}$ FWHM) and the
lines are clearly resolved. Even at 99\% confidence 
($\Delta C=9.2$ for 2 parameters), the velocity width is constrained to 
$\sigma=340\pm140$\,km\,s$^{-1}$. Upon allowing the velocity width of the
individual lines to vary, then they are constrained to lie
within the range $\sigma=250-500$\,km\,s$^{-1}$ as shown in Table\,1.  The mean outflow
velocity is $-810$\,km\,s$^{-1}$. The overall fit statistic is $C =
3804$ for 3811 bins.

We used the photoionization code \textsc{xstar} \citep{Kallman04}
to derive the parameters of the absorber, assuming the baseline
continuum described above, including the soft excess. Solar abundances
are assumed throughout \citep{GS98}.
An important input parameter is the turbulent velocity, which can effect
the absorption line equivalent widths and hence the derived column
density.  We experimented with two
different values of the turbulent velocity chosen to represent two
likely extremes: (i) a lowest value of $v_{\rm
  turb}=100$\,km\,s$^{-1}$ and (ii) $v_{\rm turb}=300$\,km\,s$^{-1}$,
the latter being consistent with the measured width of the absorption
lines. The fitted continuum parameters are $\Gamma=1.68\pm0.02$ and
for the blackbody, $kT=110\pm8$\,eV.  For case (i), then $N_{\rm H} =
(3.2\pm0.6) \times 10^{21}$\,cm$^{-2}$, the ionization parameter is
$\log \xi$\footnote{The units of $\xi$ are erg\,cm\,s$^{-1}$.}$= 2.45^{+0.13}_{-0.08}$ 
and the outflow velocity is $v_{\rm out} = -810^{+60}_{-55}$\,km\,s$^{-1}$.  The fit
statistic is $C/{\rm bins} = 3795/3811$.  For case (ii), then $N_{\rm
  H} = (1.30\pm0.25) \times 10^{21}$\,cm$^{-2}$, $\log \xi =
2.45^{+0.06}_{-0.07}$ and the outflow velocity
$v_{\rm out} = -840^{+60}_{-50}$\,km\,s$^{-1}$.  The fit statistic is
$C/{\rm bins} = 3783/3811$.  
If the warm absorber is not included in the model, then the 
fit statistic is substantially worse by $\Delta C=220$ (compared to model (ii)). 
Only a single outflowing layer of gas is required to model the warm absorber.

The higher turbulence velocity model is statistically preferred and is
consistent with the measured 340\,km\,s$^{-1}$ widths of the
lines. Either model yields an outflow velocity of $-800$\,km\,s$^{-1}$
within a statistical error of $<10$\%. 
Hereafter we adopt the parameters from model (ii), as the turbulent
velocity is consistent with widths of the individual absorption lines.
However in neither model is the fitted
column density of the absorber as high as the value of $N_{\rm H} \sim
3\times 10^{22}$\,cm$^{-2}$ reported by \citet{Torresi09b} from an
analysis of a short 34.5\,ks XMM-Newton/RGS observation of 3C\,382 on April 28, 2008.  
If the column density of the warm absorber is fixed to the value of 
$N_{\rm H} = 3\times 10^{22}$\,cm$^{-2}$ in the HETG spectrum, then the fit statistic is 
considerably worse ($C/{\rm bins} = 8390/3811$). 

As a consistency check, we analyzed 
the archival RGS data of 3C\,382 with model (ii) and the same continuum form 
as above. We found that the column density and ionization parameter were degenerate 
with each other, given the short exposure of the RGS spectrum. Thus for the RGS, 
$N_{\rm H} = 1.4^{+1.4}_{-1.3} \times 10^{22}$\,cm$^{-2}$, 
$\log \xi = 3.4\pm1.0$ and the outflow velocity
$v_{\rm out} = -1200^{+300}_{-500}$\,km\,s$^{-1}$.  
The fit statistic improves only by $\Delta C=25$ to $C/{\rm bins} =
826/820$ upon adding the absorber.
Thus within the larger errors, the parameters are consistent 
with those obtained from the HETG. The individual lines detected in the Chandra HETG 
observation 
were also compared to the absorption lines claimed by \citet{Torresi09b} on the basis 
of the RGS data. With the higher signal to noise ratio of the HETG spectrum compared to 
the RGS spectrum, 
we only confirm the detection 
of one line reported by \citet{Torresi09b}, the Fe\,\textsc{xx} line at a rest--frame 
energy of 1025\,eV\footnote{The line reported at 1.356\,keV by \citet{Torresi09b} may also be 
associated with the $1s-2p$ line of Mg\,\textsc{xi}, as noted in Table~1.} 

\section{Discussion and Conclusions}

The Chandra HETG spectra have revealed an ionized outflow in the BLRG
3C\,382. The outflow parameters are well
determined, with $N_{\rm H} = (1.30\pm0.25) \times
10^{21}$\,cm$^{-2}$, $\log \xi =
2.45^{+0.06}_{-0.07}$ and $v_{\rm out} =
-840^{+60}_{-50}$\,km\,s$^{-1}$, while the absorption line widths are
resolved with $\sigma=340\pm70$\,km\,s$^{-1}$.

To characterize the outflow, we define the (unabsorbed) ionizing
luminosity $L_{\rm ion}$, which in \textsc{xstar} is defined from
$1-1000$\,Rydberg. This depends on the continuum model fitted to the
data and it is important to take into account the soft excess.  Using
the best fit powerlaw plus blackbody continuum, then $L_{\rm ion} =
1.2\times10^{45}$\,erg\,s$^{-1}$. We note that if a different continuum form
is used to parameterize the soft excess, e.g. a broken power-law, then
the ionizing luminosity can be a factor of $\sim 2$ higher; keeping
this caveat in mind we adopt the more conservative lower luminosity
value of $1.2\times 10^{45}$\,erg\,s$^{-1}$

\subsection{The Location of the Absorber}

The upper--bound to the wind radius ($R_{\rm out}$) is determined by
geometrical constraints, i.e. if the thickness of the absorber
$\Delta R/R << 1$ (valid for a thin shell) and as $N_{\rm H} = n \Delta R$, where n is the
electron number density. As the ionization parameter of the absorber
is defined as $\xi = L_{\rm ion}/nR^{2}$, then by substitution 
$R_{\rm  out} << L_{\rm ion} / N_{\rm H} \xi = 3.3 \times 10^{21}$\,cm (or
$<<1$\,kpc)\footnote{Note the same radius is obtained by integrating 
down the line of a sight of a homogeneous radial outflow.}.
The lower bound is formed by the escape velocity,
i.e. for the gas to escape the system as an outflow then $R_{\rm esc}
> c^{2} / v^{2} R_{\rm s}$, where $R_{\rm s} = 2GM/c^2$ is the black
hole Schwarzschild radius and $v=800$\,km\,s$^{-1}$.  If for 3C\,382,
$M = 1\times10^{9} \Msun$ (with a 40\% uncertainty, see \citet{Mar04}), 
then $R_{\rm esc} > 1.4 \times 10^{5} R_{\rm s} > 4.2 \times 10^{19} {\rm cm} > 13 {\rm pc}$.  
In other words the location of the soft X-ray outflow is likely bounded between
approximately 10\,pc and 1 kpc.

Furthermore the outflow velocity and absorption line FWHM of 
$\sim 800$\,km\,s$^{-1}$ is similar to the width of the narrow optical
forbidden lines of [O\,\textsc{iii}], [O\,\textsc{i}] and [Si\,\textsc{ii}], which for 3C\,382 
lie in the range $400-600$\,km\,s$^{-1}$, as measured from the spectra of \citet{Erac94}. 
In particular the [O\,\textsc{iii}] emission line from the \citet{Erac94}
spectrum appears to have an asymmetric profile, with the blue-wing 
extending to $\sim -1070$\,km\,s$^{-1}$ from the line centroid, which appears 
$\sim370$\,km\,s$^{-1}$ broader than the red-wing. 

Although this is not evidence 
for observing outflowing gas through the direct line of sight, the [O\,\textsc{iii}] emission 
may suggest we are viewing outflowing gas along a different sight-line.
Indeed this effect is also seen in other 
radio galaxies \citep{Gelderman1994}. 
The coincidence between the soft X-ray absorbing gas and extended [O\,\textsc{iii}] 
emission has also been noted in the radio-quiet quasar, MR\,2251-178 \citep{Kaspi04}. 
Thus the origin 
of the X-ray absorption appears consistent with the optical Narrow Line Region (NLR).
The association between the soft X-ray absorption in 3C\,382 
and any rest--frame UV absorption could be tested 
with future simultaneous Chandra and HST observations.

The density of the outflow is then $n = L_{\rm ion} / \xi R^{2}$.  For
the parameters above, then $n= 0.4 - 2400$\,cm$^{-3}$. While loosely
constrained, this is consistent with typical expected NLR densities of
$\sim10^{3}$\,cm$^{-3}$ in AGN \citep{Koski78}. 
The mass outflow rate for a uniform spherical flow is $\dot{M}_{\rm out} =
4\pi n R^{2} m_{\rm p} v_{\rm out}$, where $nR^{2} = L_{\rm ion} /\xi$
and $m_p$ is the proton mass.  Hence for the measured warm absorber
parameters for 3C\,382, $\dot{M}_{\rm out} = 7.2 \times
10^{27}$\,g\,s$^{-1}$ or $\dot{M}_{\rm out}= 100\Msun$\,yr$^{-1}$.
 
However the low rate of appearance of 
such absorbers amongst BLRG's or radio-loud AGN generally might suggest 
that the solid angle subtended by individual absorbing filaments is much 
smaller than $4\pi$\,steradians. 
Therefore, the mass outflow rate could be considerably smaller than the above estimate. 
Subsequently the kinetic power of the soft X-ray outflow (for $v_{\rm
  out}=-800$\,km\,s$^{-1}$) is then $\dot{E} = 1/2 \dot{M}_{\rm out}
v_{\rm out}^{2} < 2 \times 10^{43}$\,erg\,s$^{-1}$, which
energetically is a fairly insignificant 2\% of the ionizing luminosity
and is unlikely to contribute significantly towards AGN feedback \citep{King03}.  




\acknowledgements

This research has made use of data obtained from the High Energy
Astrophysics Science Archive Research Center (HEASARC), provided by
NASA's Goddard Space Flight Center.
R.M.S. acknowledges support from
NASA through the \suzaku\ and \chandra\ programs. 
M.E. thanks the NSF for support via grant AST-0807993.
We would like to thank 
Tahir Yaqoob for assistance with the Chandra data analysis.

\begin{figure}
\begin{center}
\rotatebox{-90}{
\epsscale{0.7}
\plotone{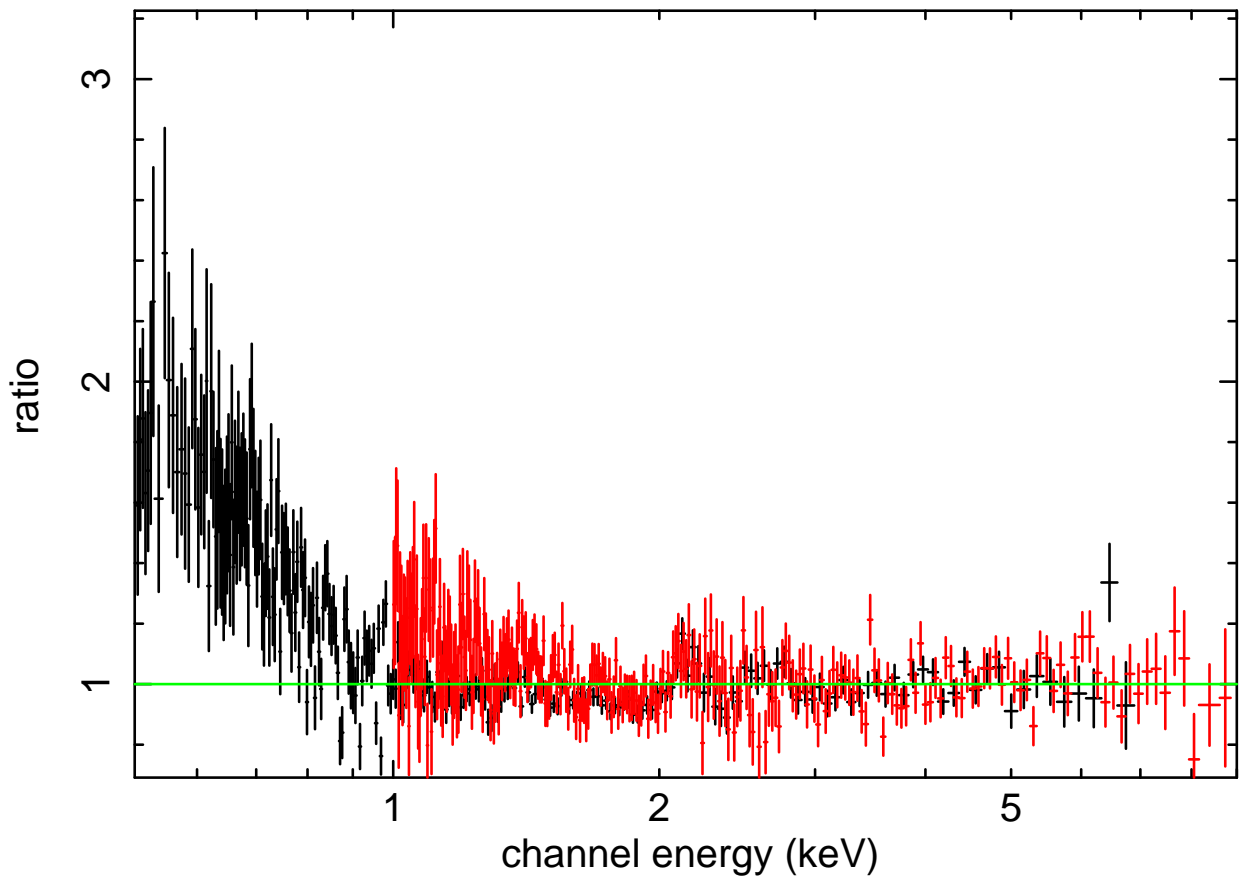}}
\caption{Chandra HETG MEG (black) and HEG (red) spectra of 3C\,382, binned 
coarsely at four times the resolution of the gratings, in order to show the broad-band 
continuum spectrum. The data is plotted as a ratio against an absorbed power-law, 
of photon index $\Gamma=1.66$. A clear excess of counts is present in the soft X-ray 
band below 1 keV, while significant residuals below unity are also present between 0.7--1.0\,keV, indicating 
that a warm absorber is present.}
\end{center}
\end{figure}

\begin{figure}
\begin{center}
\rotatebox{-90}{\includegraphics[height=12cm]{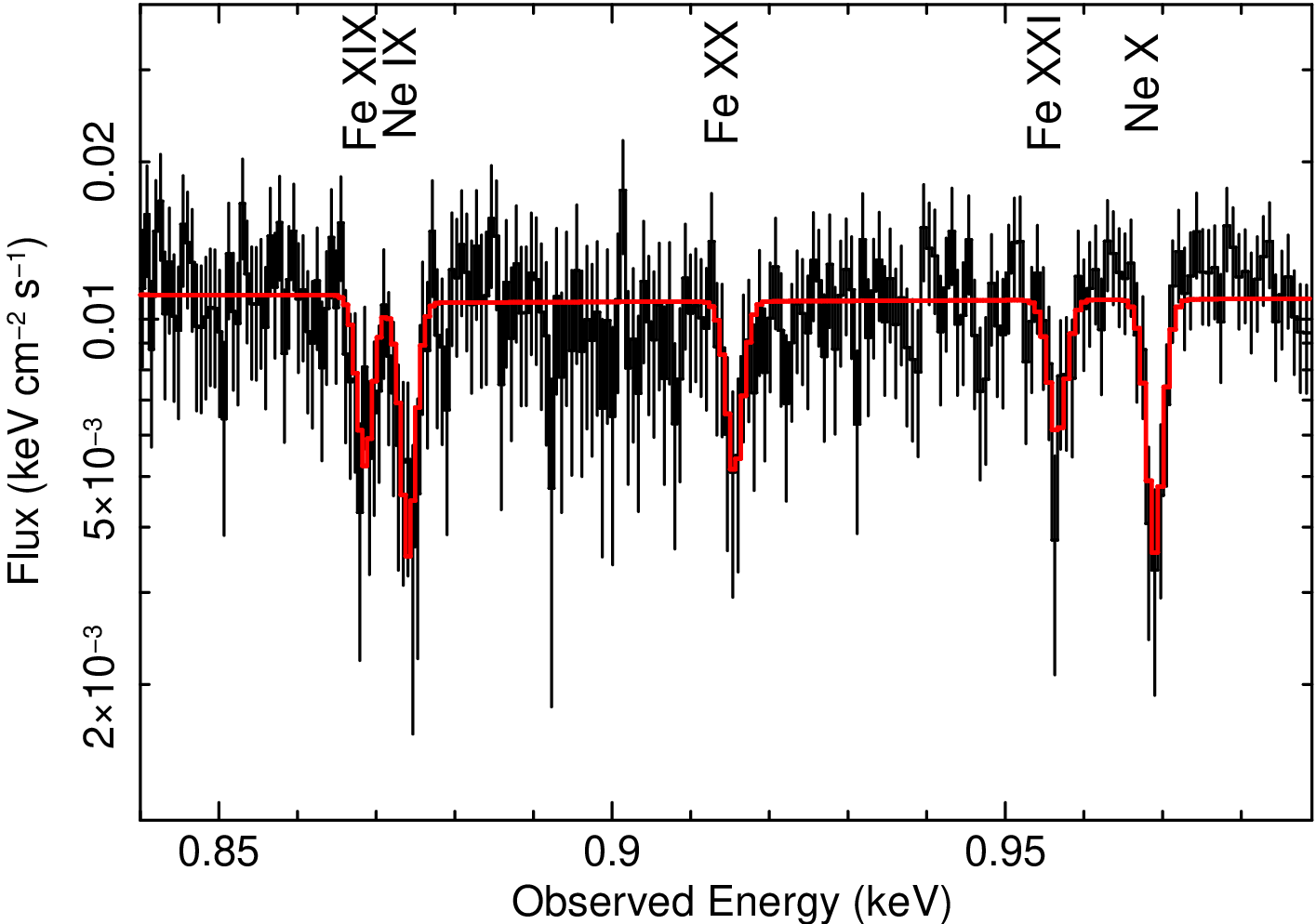}}
\rotatebox{-90}{\includegraphics[height=12cm]{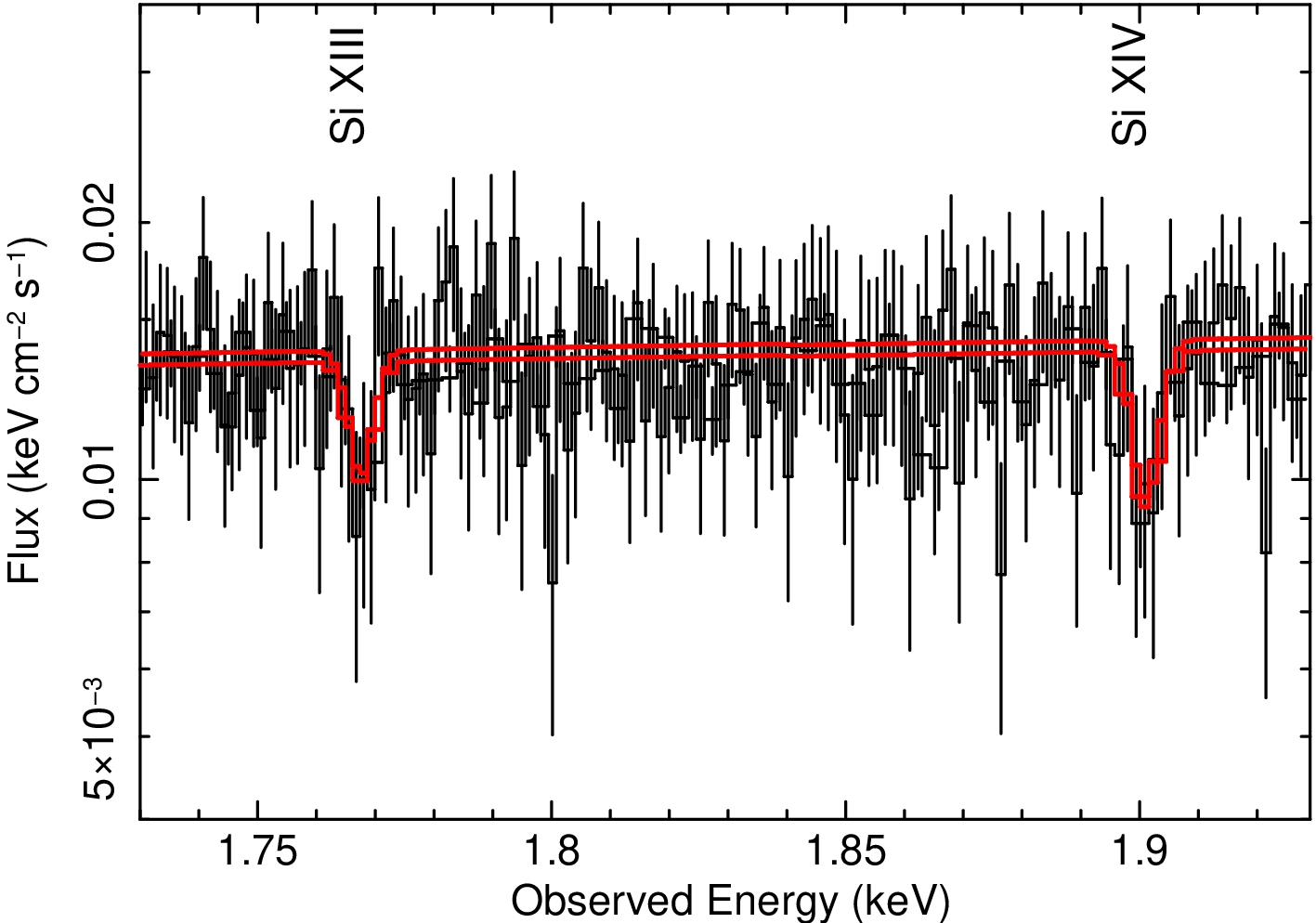}}
\end{center}
\caption{Fluxed HETG spectra showing the comparison between the data binned at 
HWHM resolution and the best-fit absorption line model (solid line) 
described in the text. Several absorption lines are present in the HETG spectrum, 
as labelled in the above figure and listed in Table 1.} 
\end{figure}



\clearpage

\begin{deluxetable}{lcccccc}
\tabletypesize{\small}
\tablecaption{Summary of HETG absorption line parameters.}
\tablewidth{0pt}
\tablehead{
\colhead{E(obs)$^{a}$} & \colhead{E(rest)$^{b}$} & \colhead{ID$^{c}$} 
& \colhead{EW$^{d}$} & \colhead{$v_{\rm out}^{e}$} & \colhead{$\sigma^{f}$} 
& \colhead{$\Delta C$$^{g}$}}

\startdata

868.5 & $918.8_{-0.9}^{+1.1}$ & Fe\,\textsc{xix} $2p-3d$ (917.0)  & $-1.4\pm0.5$ 
& $590^{+360}_{-290}$ & $509^{+470}_{-216}$ & 24 \\

874.1 & $924.7_{-0.7}^{+0.6}$ & Ne\,\textsc{ix} $1s-2p$ (922.0) & $-1.9^{+0.5}_{-0.4}$ & 
$875^{+195}_{-230}$ & $493_{-200}^{+357}$ & 35 \\
            
915.8 & $968.8_{-0.9}^{+0.5}$ & Fe\,\textsc{xx} $2p-3d$ (967.3) & $-1.6\pm0.5$ & 
$460^{+160}_{-280}$ & $516_{-400}^{+490}$ & 21 \\
    
956.6 & $1012.0_{-0.7}^{+0.7}$ & Fe\,\textsc{xxi} $2p-3d$ (1009.0) & $-1.4^{+0.4}_{-0.5}$  
& $890\pm210$ & $296_{-71}^{+400}$ & 19 \\

969.0 & $1025.1_{-0.4}^{+0.4}$ & Ne\,\textsc{x} $1s-2p$ (1021.5) & $-2.1^{+0.4}_{-0.5}$ 
& $1050\pm120$ & $289_{-102}^{+114}$ & 48 \\

1767.7  & $1870.0_{-1.3}^{+1.2}$ & Si\,\textsc{xiii} $1s-2p$ (1865.0) & $-1.6^{+0.4}_{-0.6}$ 
& $800^{+190}_{-210}$ & $255_{-165}^{+272}$ & 26 \\

1901.1 & $2011.1_{-1.2}^{+0.3}$ & Si\,\textsc{xiv} $1s-2p$ (2004.4) & $-2.2\pm0.7$ & 
$1000^{+50}_{-180}$ & $284_{-180}^{+308}$ & 32 \\

1281.4 & $1355.6_{-1.0}^{+1.0}$ & Mg\,\textsc{xi} $1s-2p$ (1352.2) & $-1.0\pm0.4$ & 
$750\pm220$ & $340^{h}$ & 12 \\

1395.4 & $1476.2_{-1.0}^{+1.0}$ & Mg\,\textsc{xii} $1s-2p$ (1472.2) & $-0.8\pm0.4$ & 
$810\pm203$ & $340^{h}$ & 9 \\

\enddata


\tablenotetext{a}{Observed energy of absorption line in eV.} 
\tablenotetext{b}{Energy of absorption line in rest-frame of 3C\,382, in units eV.} 
\tablenotetext{c}{Line identification and lab frame energy in eV in parenthesis. Atomic data 
are from http://physics.nist.gov}
\tablenotetext{d}{Equivalent width, units eV.}
\tablenotetext{e}{Outflow velocity of absorption line, in units km\,s$^{-1}$.}
\tablenotetext{f}{$1\sigma$ velocity width of absorption line, in units km\,s$^{-1}$.}
\tablenotetext{g}{Improvement in C-statistic, upon adding line to model.}
\tablenotetext{h}{Parameter is fixed in the model.}

\end{deluxetable}

\clearpage

\end{document}